\documentclass[12pt]{article}

\usepackage{amsmath,amssymb}
\usepackage{bm}
\usepackage{graphicx, color}
\usepackage{wrapfig}
\usepackage{cite}
\begin{document}
\title{
\begin{flushright}
\ \\*[-80pt] 
\begin{minipage}{0.2\linewidth}
\normalsize
EPHOU-18-017 \\
HUPD1810 \\*[50pt]
\end{minipage}
\end{flushright}
{\Large \bf 
Finite modular subgroups for fermion mass matrices\\*[10pt]
 and baryon/lepton number violation
\\*[20pt]}}

\author{ 
\centerline{
~Tatsuo Kobayashi $^{1}$,
~~Yusuke~Shimizu $^{2}$,~~~~Kenta~Takagi $^{2}$,} \\*[5pt]
\centerline{~~Morimitsu Tanimoto $^{3}$, 
~~Takuya H. Tatsuishi $^{1}$,~~Hikaru Uchida $^{1}$} 
\\*[20pt]
\centerline{
\begin{minipage}{\linewidth}
\begin{center}
$^1${\it \normalsize
Department of Physics, Hokkaido University, Sapporo 060-0810, Japan} \\*[5pt]
$^2${\it \normalsize
Graduate School of Science, Hiroshima University, Higashi-Hiroshima 739-8526, Japan} \\*[5pt]
$^3${\it \normalsize
Department of Physics, Niigata University, Niigata 950-2181, Japan}
\end{center}
\end{minipage}}
\\*[50pt]}

\date{
\centerline{\small \bf Abstract}
\begin{minipage}{0.9\linewidth}
\medskip 
\medskip 
\small
We study a flavor model that the quark sector has the $S_3$ modular symmetry while the lepton sector has the $A_4$ modular symmetry. Our model leads to characteristic quark mass matrices which are consistent with experimental data of quark masses, mixing angles and the CP violating phase. The lepton sector is also consistent with the experimental data of neutrino oscillations. We also study baryon and lepton number violations in our flavor model. 
\end{minipage}
}

\begin{titlepage}
\maketitle
\thispagestyle{empty}
\end{titlepage}

\section{Introduction}

The standard model is now well established by the discovery of the Higgs boson.
However, the flavor puzzle of quarks and leptons is still an open question.
In order to understand the origin of the flavor structure, considerable interests in discrete flavor symmetries
\cite{Altarelli:2010gt,Ishimori:2010au,Ishimori:2012zz,Hernandez:2012ra,King:2013eh,King:2014nza,Tanimoto:2015nfa,King:2017guk,Petcov:2017ggy}
have been promoted by early works of the quark masses and mixing angles 
\cite{Pakvasa:1977in,Wilczek:1977uh}
and the recent developments in the neutrino oscillation experiments.
Indeed, the neutrino oscillation experiments have determined the two neutrino mass squared differences and three neutrino mixing angles precisely.
In particular, the atmospheric neutrino mixing angle $\theta_{\rm atm}$ has been observed near the maximal angle $45^\circ$ \cite{Tanabashi:2018oca}.
This may suggest a flavor symmetry, e.g. $\mu$-$\tau$ symmetry.

Many models have been proposed by using $S_3$, $A_4$, $S_4$, $A_5$ and other groups with larger orders to explain large neutrino mixing angles,
in which symmetry breakings are required to reproduce realistic mixing angles
\cite{Petcov:2018snn}.
The effective Lagrangian of a typical flavor model is given by introduced gauge singlet scalars which are so-called flavons.
Their vacuum expectation values (VEVs) determine the flavor structure of quarks and leptons.
A complicated vacuum alignment is frequently required to reproduce the experimental data in a model with many flavons .

Superstring theory with certain compactifications can lead to non-Abelian discrete flavor symmetries.
For example, heterotic orbifold models lead to $D_4$, $\Delta(54)$, etc. \cite{Kobayashi:2006wq}.
(See also \cite{Kobayashi:2004ya,Ko:2007dz,Beye:2014nxa,Olguin-Trejo:2018wpw,Nilles:2018wex}.)
Similar flavor symmetries are also derived in type II magnetized and intersecting D-brane models 
\cite{Abe:2009vi,Abe:2009uz,BerasaluceGonzalez:2012vb,Marchesano:2013ega,Abe:2014nla}.
On the other hand, string theory on tori or orbifolds has the modular symmetry which 
acts non-trivially on flavors of quarks and leptons
\cite{Lauer:1989ax,Lerche:1989cs,Ferrara:1989qb,Cremades:2004wa,Kobayashi:2017dyu,Kobayashi:2018rad}.
In this sense, the modular symmetry is a non-Abelian discrete flavor symmetry.

It is interesting that the modular group $\Gamma$ includes $S_3$, $A_4$, $S_4$, and $A_5$
as its finite subgroups \cite{deAdelhartToorop:2011re}.
Remarkably, there is a significant difference between a model based on a finite modular symmetry and that based on a usual non-Abelian discrete flavor symmetry.
In a theory based on a finite modular symmetry, Yukawa couplings are written in terms of modular forms: holomorphic functions of a complex scalar field, the modulus $\tau$; and they transform non-trivially under the modular symmetry.

In this aspect, an attractive ansatz was proposed by taking $\Gamma_3 \simeq A_4$ in Ref.\cite{Feruglio:2017spp},
where both left-handed leptons and right-handed neutrinos of $A_4$ triplets while Yukawa couplings are $A_4$ triplets of modular forms.
Along with this work, models of $\Gamma_2 \simeq S_3$ \cite{Kobayashi:2018vbk},
$\Gamma_4 \simeq S_4$ \cite{Penedo:2018nmg} and $\Gamma_5 \simeq A_5$ \cite{Novichkov:2018nkm} have been proposed.
The numerical discussions of the neutrino flavor mixing have been done based on $A_4$ \cite{Criado:2018thu,Kobayashi:2018scp,Novichkov:2018yse} and $S_4$ \cite{Novichkov:2018ovf} modular groups respectively.
In particular, a comprehensive analysis of the $A_4$ modular group have provided a clear prediction of the neutrino mixing angle and the CP violating phase \cite{Kobayashi:2018scp}.
Moreover, the $A_4$ modular symmetry has been applied to the $SU(5)$ grand unified theory of leptons and quarks \cite{deAnda:2018ecu}.
Furthermore, modular forms for $\Delta(96)$ and $\Delta(384)$ were constructed \cite{Kobayashi:2018bff}.

Our purpose is to study the quark mixing angles and the CP violating phase,
which were a main target of the early challenge for flavors \cite{Pakvasa:1977in,Wilczek:1977uh}.
Superstring theory has six-dimensional compact space $X^6$ in addition to our four-dimensional spacetime.
Suppose that the six-dimensional compact space is a product of three two-dimensional compact spaces, 
$X^6 = X^2_1\times X^2_2 \times X^2_3$.
Note that each $X^2_i$ can have geometrical symmetry such as the modular symmetry.
Flavor differences can originate from one of the two-dimensional compact spaces $X^2_i$, while moduli of the other $X^2_j$'s $(j\neq i)$ can contribute not to ratios of the Yukawa couplings but to an overall factor of them.
For example, three generations (quasi-)localize at different points on one of $X^2_i$'s \footnote{
Alternatively, in heterotic orbifold models, two generations localize at two fixed points on the two-dimensional 
orbifold, but one generation is a bulk mode.(See e.g. \cite{Kobayashi:2004ya}.)}
while they localize at the same point on the other $X^2_i$'s.
If the flavor difference of quarks and leptons is originated from a same two-dimensional compact space, 
the quarks and leptons have same flavor symmetry and the same value of $\tau$.
For example, if the lepton sector has the $A_4$ modular symmetry, 
the same symmetry controls Yukawa couplings in the quark sector.
Indeed, such a setup is required by the four-dimensional grand unified theory.
However, the hierarchical mass structure and small mixing angles of quarks are remarkably distinguished from those of leptons.
If the up-quark and down-quark mass matrices have the same structure as the charged lepton mass matrix in Ref.\cite{Kobayashi:2018scp},
it is very difficult to reproduce the observed hierarchical three CKM mixing angles and the CP violating phase.
It is because the constraints from the precise observed mixing angles and the CP violating phase are strong. 

In this paper, we make an alternative ansatz.
The flavor difference in the quark sector originates from a two-dimensional compact space while that in the lepton sector originates from another two-dimensional compact space.
Thus, we assume two different flavor symmetries, i.e. two different finite modular subgroups for the quark and lepton sectors.
The modulus parameters for each sector are defined independently.
In our present model, we assume the $S_3$ modular symmetry for the quark sector and the $A_4$ modular symmetry for the lepton sector.
We use the same lepton sector as given in Ref.\cite{Kobayashi:2018scp}.
Since the irreducible representations of $ S_3$ group are $\bf 2$, $\bf 1$ and $\bf 1'$,
we have an advantage to explain the hierarchy of three quark mixing angles.
Based on a numerical analysis, we present realistic quark mass matrices in this work.
Furthermore, we can discuss the violation of baryon number and lepton number
in the case that the lepton sector is still controlled by the finite modular symmetry
$\Gamma_3 \simeq A_4$.
Then, we find interesting results.


The paper is organized as follows.
In section 2, we give a brief review on modular symmetry of $A_4$ and $S_3$. 
In section 3, we summarize the result of our previous work of leptons in the finite modular symmetry $\Gamma_3 \simeq A_4$.
In section 4, we present a quark mass matrix and their numerical results under the finite modular symmetry $\Gamma_2 \simeq S_3$.
In section 5, the baryon and lepton number violations are discussed.
Section 6 is devoted to a summary.
Appendix A shows the relevant multiplication rules of $A_4$. 
Also, Appendix B shows relevant tensor products of $S_3$ modular forms.

\section{ $A_4$ and $S_3$ in modular group}

We briefly review the modular symmetry on a torus and its low-energy effective field theory.
The torus compactification is the simplest compactification.
For example, the two-dimensional torus $T^2$ can be constructed as division of $\mathbb{R}^2$ by a two-dimensional lattice $\Lambda$, i.e. $T^2=\mathbb{R}^2/\Lambda$.
Here, we use the complex coordinate on $\mathbb{R}^2$ and the lattice is spanned by two lattice vectors, $\alpha_1=2 \pi R$ and $\alpha_2 = 2 \pi R \tau$; where $R$ is real and $\tau$ is a complex modulus parameter.
There is ambiguity in choice of the basis vectors.
The same lattice can be spanned by the following basis vectors,
\begin{equation}
	\label{eq:SL2Z}
	\left(
	\begin{array}{c}
		\alpha'_2 \\ \alpha'_1
	\end{array}
	\right) =\left(
	\begin{array}{cc}
		a & b \\
		c & d
	\end{array}
	\right) \left(
	\begin{array}{c}
		\alpha_2 \\ \alpha_1
	\end{array}
	\right) \ ,
\end{equation}
where $a,b,c,d$ are integer with satisfying $ad-bc = 1$.
That is the $SL(2,\mathbb{Z})$ transformation.
Under the above transformation, the modulus parameter $\tau \equiv \alpha_2/\alpha_1$ transforms as 
\begin{equation}\label{eq:tau-SL2Z}
	\tau \longrightarrow \tau' = \gamma\tau= \frac{a\tau + b}{c \tau + d}\ ,
\end{equation}
and this modular transformation is generated by $S$ and $T$: 
\begin{eqnarray}
	& &S:\tau \longrightarrow -\frac{1}{\tau}\ , \\
	& &T:\tau \longrightarrow \tau + 1\ .
\end{eqnarray}
They satisfy the following algebraic relations, 
\begin{equation}
	S^2 =\mathbb{I}\ , \qquad (ST)^3 =\mathbb{I}\ .
\end{equation}
If we impose $T^N=\mathbb{I}$ furthermore,
we obtain finite subgroups $\Gamma_N$, and $\Gamma_N$ with $N=2,3,4,5$ are isomorphic to
$S_3$, $A_4$, $S_4$ and $A_5$, respectively \cite{deAdelhartToorop:2011re}.
Indeed, $\Gamma_N$ is a quotient of the modular group by the so-called congruence subgroup ${\Gamma}(N)$.
Holomorphic functions which transform as
\begin{equation}
	f(\tau)\to (c\tau+d)^kf(\tau)~,
\end{equation}
under the modular transformation Eq.(\ref{eq:tau-SL2Z}) are called modular forms of weight $k$.

Superstring theory on the torus $T^2$ or orbifold $T^2/Z_N$ has the modular symmetry.
Its low-energy effective field theory is described in terms of supergravity theory,
and the string-derived supergravity theory has also the modular symmetry.
Under the modular transformation (\ref{eq:tau-SL2Z}), chiral superfields $\phi^{(I)}$ transform as \cite{Ferrara:1989bc},
\begin{equation}
\label{eq:modular-chiral}
	\phi^{(I)}\to(c\tau+d)^{-k_I}\rho^{(I)}(\gamma)\phi^{(I)},
\end{equation}
where $-k_I$ is the so-called modular weight and $\rho^{(I)}(\gamma)$ denotes a unitary representation matrix of $\gamma\in\Gamma_N$.
The kinetic terms of their scalar components are written by 
\begin{equation}
	\sum_I\frac{|\partial_\mu\phi^{(I)}|^2}{\langle-i\tau+i\bar{\tau}\rangle^{k_I}} ~,
	\label{kinetic}
\end{equation}
which is invariant under the modular transformation.
Here, we use the convention that the superfield and its scalar component are denoted by the same letter.
The superpotential should be also invariant under the modular symmetry.
In other words, the superpotential should have vanishing modular weight in global supersymmetric models.
On the other hand, the superpotential in supergravity should be invariant under the modular symmetry up to the K\"ahler transformation.
We study the minimal supersymmetric standard model (MSSM), one of global supersymmetric models,
and its extension with right-handed neutrinos in the following sections.
Thus, the superpotential of our model has vanishing modular weight.
We note that Yukawa couplings as well as higher order couplings depend on modulus and can have non-vanishing modular weights.
The breaking scale of supersymmetry can be between $\mathcal{O}(1)$TeV and the compactification scale.
The modular symmetry is broken by the vacuum expectation value of $\tau$ at the compactification scale which is the Planck scale or slightly lower scale order.

For $\Gamma_3 \simeq A_4$, the dimension of the linear space ${\cal M}_k(\Gamma_3)$ of modular forms of weight $k$ is $k+1$.
In other words, there are three linearly independent modular forms of the lowest non-trivial weight $2$
\cite{Feruglio:2017spp,Gunning:1962,Schoeneberg:1974,Koblitz:1984}.
These forms have been explicitly obtained \cite{Feruglio:2017spp} in terms of the Dedekind eta-function $\eta(\tau)$: 
\begin{equation}
	\eta(\tau) = q^{1/24} \prod_{n =1}^\infty (1-q^n)~,
\end{equation}
where $q = e^{2 \pi i \tau}$ and $\eta(\tau)$ is a modular form of weight~$1/2$.
In what follows, we use the following basis of the $A_4$ generators $S$ and $T$ in the triplet representation:
\begin{align}
	\begin{aligned}
	S=\frac{1}{3}
	\begin{pmatrix}
		-1 & 2 & 2 \\
		2 &-1 & 2 \\
		2 & 2 &-1
	\end{pmatrix},
	\end{aligned}
	\qquad \qquad
	\begin{aligned}
	T=
	\begin{pmatrix}
		1 & 0& 0 \\
		0 &\omega& 0 \\
		0 & 0 & \omega^2
	\end{pmatrix}, 
	\end{aligned}
\end{align}
where $\omega=e^{2\pi i/3}$ .
The modular forms transforming as a triplet of $A_4$ can be written in terms of 
$\eta(\tau)$ and its derivative \cite{Feruglio:2017spp}:
\begin{eqnarray} 
	\label{eq:Y-A4}
	Y_1^{A_4}(\tau) &=& \frac{i}{2\pi}\left( \frac{\eta'(\tau/3)}{\eta(\tau/3)} +\frac{\eta'((\tau +1)/3)}{\eta((\tau+1)/3)}
	+\frac{\eta'((\tau +2)/3)}{\eta((\tau+2)/3)} - \frac{27\eta'(3\tau)}{\eta(3\tau)} \right), \nonumber \\
	Y_2^{A_4}(\tau) &=& \frac{-i}{\pi}\left( \frac{\eta'(\tau/3)}{\eta(\tau/3)}  +\omega^2\frac{\eta'((\tau +1)/3)}{\eta((\tau+1)/3)}  
	+\omega \frac{\eta'((\tau +2)/3)}{\eta((\tau+2)/3)}  \right) , \label{tripletY} \\ 
	Y_3^{A_4}(\tau) &=& \frac{-i}{\pi}\left( \frac{\eta'(\tau/3)}{\eta(\tau/3)}  +\omega\frac{\eta'((\tau +1)/3)}{\eta((\tau+1)/3)}  
	+\omega^2 \frac{\eta'((\tau +2)/3)}{\eta((\tau+2)/3)}  \right)\,.
	\nonumber
	\label{YA4}
\end{eqnarray}
These are expressed as the following $q$-expansions:
\begin{align}
	Y^{A_4}=\begin{pmatrix}Y_1^{A_4}(\tau)\\
	Y_2^{A_4}(\tau)\\
	Y_3^{A_4}(\tau)\end{pmatrix}=
	\begin{pmatrix}
	1+12q+36q^2+12q^3+\dots \\
	-6q^{1/3}(1+7q+8q^2+\dots) \\
	-18q^{2/3}(1+2q+5q^2+\dots)\end{pmatrix}.
\end{align}
They satisfy the constraint \cite{Feruglio:2017spp}:
\begin{align}
	(Y_2^{A_4}(\tau))^2+2Y_1^{A_4}(\tau) Y_3^{A_4}(\tau)=0~.
	\label{condition}
\end{align}

For $\Gamma_2 \simeq S_3$, the dimension of the linear space ${\cal M}_k(\Gamma_2)$ of modular forms of weight $k$ is $k/2+1$. 
In other words, there are two linearly independent modular forms of the lowest non-trivial weight $2$.
The $S_3$ doublet modular forms of weight 2 are presented in Ref. \cite{Kobayashi:2018vbk},
\begin{eqnarray} 
	\label{eq:Y-S3}
	Y_{1}^{S_3}(\tau) &=& \frac{i}{4\pi}\left( \frac{\eta'(\tau/2)}{\eta(\tau/2)}  +\frac{\eta'((\tau +1)/2)}{\eta((\tau+1)/2)}  
	- \frac{8\eta'(2\tau)}{\eta(2\tau)}  \right), \nonumber \\
	Y_2^{S_3}(\tau) &=& \frac{\sqrt{3}i}{4\pi}\left( \frac{\eta'(\tau/2)}{\eta(\tau/2)}  -\frac{\eta'((\tau +1)/2)}{\eta((\tau+1)/2)}   \right) , \label{doubletY}  \nonumber
\end{eqnarray}
where we use the following basis of $S_3$ generators $S$ and $T$ in the doublet representation:
\begin{equation}
	S = \frac{1}{2}\left(
	\begin{array}{cc}
		-1 & -\sqrt{3} \\
		-\sqrt{3} & 1
	\end{array}\right), \qquad\qquad 
	T = \left(
	\begin{array}{cc}
		1 & 0 \\
		0 & -1
	\end{array}\right).
	\label{S3base}
\end{equation}
The doublet modular forms have the following  $q$-expansions:
\begin{align}
	Y^{S_3}=\begin{pmatrix}Y_1^{S_3}(\tau)\\Y_2^{S_3}(\tau)
	\end{pmatrix}=
	\begin{pmatrix}
	\frac{1}{8}+3q+3q^2+12q^3+3q^4+\dots \\
	\sqrt{3}q^{1/2}(1+4q+6q^2+8q^3+\dots ) \end{pmatrix}.
	\label{expansionS3}
\end{align}

In order to present realistic quark mass matrices, we discuss modular forms of weight $k$, $Y^{S_3(k)}$.
Since we work in the basis of Eq.(\ref{S3base}), the tensor product of two doublets is expanded by 
\begin{eqnarray}
	\left(
	\begin{array}{c}
		x_1 \\ x_2 
	\end{array}\right)_{\bf 2} \otimes
	\left(
	\begin{array}{c}
		y_1 \\ y_2 
	\end{array}\right)_{\bf 2} &=&\left(x_1y_1+x_2y_2\right)_{\bf 1} 
	\oplus\left(x_1y_2-x_2y_1\right)_{\bf 1'}
	\oplus\left(
	\begin{array}{c}
		x_1 y_1-x_2 y_2 \\ -x_1 y_2- x_2 y_1 
	\end{array}\right)_{\bf 2}.
	\label{tensorproduct}
\end{eqnarray}
By using the tensor product of the two doublets $(Y^{S_3}_1(\tau),Y^{S_3}_2(\tau))^T$, we can construct modular forms of weight 4,  $Y^{{S_3}(4)}$.
The $S_3$ singlet ${\bf 1}$ modular form of weight 4 is written by 
\begin{equation} 
	{\bf 1}~:~Y^{S_3}_1(\tau)^2+Y^{S_3}_2(\tau)^2 ~ ,
\end{equation}
while the $S_3$ doublet modular forms of weight 4 is written by 
\begin{align}
	\begin{aligned}
	{\bf 2}~: ~~ Y^{{S_3}(4)}=
	\begin{pmatrix}
	Y^{S_3}_1(\tau)^2 - Y^{S_3}_2(\tau)^2  \\
	-2Y^{S_3}_1(\tau)Y^{S_3}_2(\tau) 
	\end{pmatrix}~ .
	\end{aligned}
\end{align}
On the other hand, the $S_3$ singlet ${\bf 1}'$ modular form of the weight 4 vanishes.
In conclusion, we have found 3 modular forms in the case of weight 4.
It is understandable because the number of modular forms of weight $k$ is $k/2+1$. 
 
\section{Lepton mass matrices in $A_4$ modular symmetry} 

The $A_4$ flavor model has been discussed in the lepton sector by introducing flavons
\cite{Ma:2001dn,Babu:2002dz,Altarelli:2005yp,Altarelli:2005yx,Shimizu:2011xg,Kang:2018txu}.
On the other hand, a modular invariant flavor model with the $A_4$ symmetry can explain the large mixing angles of lepton flavors without flavons.
We have already obtained successful result of the lepton sector in $A_4$ modular symmetry \cite{Kobayashi:2018scp}.
In order to clarify the difference in the flavor structure of mass matrices between the quarks and leptons,
we briefly summarize our previous results of the lepton sector and add discussions of the feature of our lepton model.

We suppose that three left-handed lepton doublets $L_i$ are compiled in a triplet of $A_4$.
The three right-handed neutrinos $\nu_i$ are also of a triplet of $A_4$.
On the other hand, the Higgs doublets, $H_{u,d}$ are supposed to be singlets of $A_4$. 
We assign three right-handed charged leptons for three different singlets of $A_4$ as $(e_1,e_2,e_3)=(e,\mu,\tau)=(1,1'',1')$.
Therefore, there are three independent couplings $\alpha$, $\beta$
and $\gamma$,  in the superpotential of the charged lepton sector.
Those coupling constants can be adjusted to the observed charged lepton masses.
The assignments of representations and modular weights to the MSSM fields and right-handed neutrino superfields are presented in Table \ref{tb:fields0}.

\begin{table}[h]
	\centering
	\begin{tabular}{|c||c|c|c|c|c|} \hline 
		&$L$&$e,~\mu,~\tau$&$\nu$&$H_u$&$H_d$\\ \hline \hline 
		\rule[14pt]{0pt}{0pt}
		$SU(2)$&$2$&$1$&$1$&$2$&$2$\\
		$A_4$&$3$& $1$,\ $1''$,\ $1'$&$3$&$1$&$1$\\
		$-k_I$&$-1$&$-1 $&$-1$&0&0 \\ \hline
	\end{tabular}
	\caption{
		The charge assignment of $SU(2)$, $A_4$, and the modular weight $-k_I$.
		The right-handed charged leptons are assigned three different $A_4$ singlets, respectively.}
	\label{tb:fields0}
\end{table}

The modular invariant mass terms of the leptons are given as the following superpotentials:
\begin{align}
	W_e&=\alpha eH_d(LY^{A_4})_{\bf 1}+\beta \mu H_d(LY^{A_4})_{\bf 1'}+
	\gamma \tau H_d(LY^{A_4})_{\bf 1"}~,\label{charged} \\
	W_D&=g(\nu H_u L Y^{A_4})_{\bf 1}~,  \label{Dirac}\\
	W_N&=\Lambda(\nu \nu Y^{A_4})_{\bf 1}~, \label{Majorana}
\end{align}
where sums of the modular weights vanish.
The parameters $\alpha$, $\beta$, $\gamma$, $g$, and $\Lambda$ are constant coefficients.
The functions $Y_i^{A_4}(\tau)$ are $A_4$ triplet modular forms of weight $2$ in Eq.(\ref{YA4}).

The superpotential of Eq.(\ref{charged}) leads to the following charged leptons mass matrix:
\begin{align}
\begin{aligned}
	M_E&=v_d~ {\rm diag}[\alpha, \beta, \gamma]
	\begin{pmatrix}
	Y_1 & Y_3 & Y_2 \\
	Y_2 & Y_1 & Y_3 \\
	Y_3 & Y_2 & Y_1
	\end{pmatrix}_{RL},
\end{aligned}\label{eq:CL}
\end{align}
where  $v_d=\langle H_d \rangle $ and we omit the superscript $A_4$ of $Y_i^{A_4}$ hereafter.
The coefficients $\alpha$, $\beta$, and $\gamma$ 
are taken to be real positive by rephasing  right-handed charged lepton fields
without loss of generality.
Those parameters can be written in terms of the modulus parameter $\tau$ and the charged lepton masses.
The superpotential of Eq.(\ref{Dirac}) gives the Dirac neutrino mass matrix as:
\begin{align}
	M_D=v_u\begin{pmatrix}
	2g_1Y_1 & (-g_1+g_2)Y_3 & (-g_1-g_2)Y_2 \\
	(-g_1-g_2)Y_3 & 2g_1Y_2 & (-g_1+g_2)Y_1 \\
	(-g_1+g_2)Y_2 & (-g_1-g_2)Y_1 & 2g_1Y_3\end{pmatrix}_{RL},
	\label{MD}
\end{align}
where $v_u=\langle H_u \rangle $.
On the other hand, the right-handed Majorana neutrino mass matrix is obtained from the superpotential of Eq.(\ref{Majorana}):
\begin{align}
	M_N=\Lambda\begin{pmatrix}
	2Y_1 & -Y_3 & -Y_2 \\
	-Y_3 & 2Y_2 & -Y_1 \\
	-Y_2 & -Y_1 & 2Y_3\end{pmatrix}_{RR}.
	\label{MajoranaR}
\end{align}
Finally, the effective neutrino mass matrix is obtained through the type I seesaw as follows:
\begin{align}
\label{eq:seesaw-mass}
	M_\nu=-M_D^{\rm T}M_N^{-1}M_D ~.
\end{align}

By fixing the modulus $\tau$ of $A_4$, the modular invariance is broken and the lepton mass matrices give the mass eigenvalues and flavor mixing numerically.
We fix $\tau$ by taking account of experimental data of NuFIT 3.2 with the $3\,\sigma$ range \cite{NuFIT}
\footnote{
We have neglected the corrections by the renormalization although the numerical analysis should be presented at a high energy scale (GUT scale) in principle.
Indeed, the quantum corrections were estimated numerically, for example, in \cite{Haba:1999fk} where the corrections are very small as far as the neutrino mass scale is smaller than $200$\,meV and $\tan \beta \leq 10$ is taken. See also \cite{Criado:2018thu}.}.

The coefficients $\alpha/\gamma$ and $\beta/\gamma$ in the charged lepton mass matrix are given only in terms of $\tau$ after inputting the observed values $m_e/m_\tau$ and $m_\mu/m_\tau$.
Indeed, $\alpha/\gamma$ and $\beta/\gamma$ are hierarchical in order to reproduce the observed charged lepton masses.
Then, we have two free parameters, $g_1/g_2$ and the modulus $\tau$ apart from the overall factors in the neutrino sector.

By inputting the data of $\Delta m_{\rm atm}^2 \equiv m_3^2-m_1^2$, $\Delta m_{\rm sol}^2 \equiv  m_2^2-m_1^2$,
and the three mixing angles $\theta_{23}$, $\theta_{12}$, and $\theta_{13}$ within $3\,\sigma$ range,
we fix completely the modulus $\tau$ and $g_1/g_2$.
Then, we can predict the CP violating Dirac phase $\delta_{CP}$ and two Majorana phases

We obtained successful results of the mixing angles for the normal hierarchy (NH) of neutrino masses $m_1<m_2<m_3$ where $m_1$, $m_2$, and $m_3$ denote three light neutrino mass eigenvalues.
The predicted Dirac CP violating phase $\delta_{CP}$ depends significantly on $\sin^2\theta_{23}$.
It is emphasized that $\sin^2\theta_{23}$ is restricted to be larger than $0.54$, and $\delta_{CP} = \pm (50^{\circ}\mbox{--}180^{\circ})$.
Since the correlation of $\sin^2\theta_{23}$ and $\delta_{CP}$ is characteristic, this prediction is testable in the future experiments of neutrinos.

We also obtained a prediction of the effective mass $m_{ee}$ which is the measure of the neutrinoless double beta decay.
It is remarkable that $m_{ee}$ is around $22$\,meV, which is testable in the future experiments of the neutrinoless double beta decay.

The obtained neutrino masses indicate nearly degenerate neutrino mass spectrum.
The sum of neutrino masses is predicted around $145$meV, which corresponds to $m_1\simeq m_2\simeq 40$\,meV and $m_3\simeq 65$\,meV.
This prediction is compared with the cosmological observations of Planck 2018 which provide us its cosmological upper bound for sum of neutrino masses: $120$ -- $160$~meV at the $95\%$ C.L. depending on the combined data \cite{Aghanim:2018eyx,Giusarma:2016phn,Vagnozzi:2017ovm}.
On the other hand, our lepton mass matrices are ruled out for the inverted hierarchy of neutrino masses $m_3 < m_1 \leq m_2$
since the sum of neutrino masses exceeds $200$meV.
Therefore, the sum of neutrino masses provides us a crucial test of our modular invariant flavor model with the $A_4$ symmetry in the near future since our model does not allow the sum of neutrino masses less than $145$meV.
 
Our modular invariant flavor model with the $A_4$ symmetry gives the large mixing angles of lepton flavors.
However, it is very difficult to reproduce observed three CKM mixing angles \cite{Okada:2018yrn}
since the observed quark mixing angles are considerably hierarchical.
Therefore, we consider another finite modular group $\Gamma_2 \simeq S_3$ for the quark sector.


\section{ CKM mixing  in $S_3$ modular symmetry}
\subsection{ Quark mass matrices }
Let us construct a quark mass matrix by using the modular forms of $S_3$ shown in the previous section.
Since the modulus parameters of $S_3$ and $A_4$ are different, we express the modulus in $S_3$ as $\tau'$.

We assign the third family of the left-handed quark $Q_3$ and right-handed quarks $u_3$ and $d_3$ to the $S_3$ non-trivial singlet ${\bf 1}'$.
The first and second families of the left-handed quarks, $Q_1$ and $Q_2$, and the right-handed quarks, $u_1$, $d_1$, $u_2$,
and $d_2$ are the $S_3$ doublet so that the first and second families correspond to the first and second components in their doublet respectively.
Furthermore, we assume that the first, second and third families have the modular weights, $-2$, $-2$, $0$ respectively.
The superpotential has a vanishing modular weight.
These assignments are summarized in Table {\ref{tb:fields} where $q_i$ denotes the right-handed up-type or down-type quarks, $u_i$ or $d_i$, for the three families ($i = 1,2,3$).
	
We can also study the model that the third family is the $S_3$ trivial singlet while the first and second families correspond to the second and first elements of $S_3$ doublet, respectively.
This model leads to almost same results as our present model.

\begin{table}[h]
	\centering
	\begin{tabular}{|c||c|c|c|c|c|c|} \hline	
		&$(Q_{1},Q_{2})$ &$Q_{3}$& $(q_{1},q_{2})$& $q_3$&$H_u$&$H_d$\\ \hline\hline 
		\rule[14pt]{1pt}{1pt}
		$SU(2)$&$\bf 2$ & $\bf 2$ & $\bf 1$ &$\bf 1$&$\bf 2$&$\bf 2$\\
		$S_3$&$\bf 2$& $\bf 1'$&  $\bf 2$& $\bf 1'$ &$\bf 1$&$\bf 1$\\
		$-k_I$&$-2$& $0$&$-2$ &$0$ &0&0 \\ \hline
	\end{tabular}
	\caption{
		The assignments of $S_3$ representations and	modular weights $-k_I$ to the MSSM fields.}
	\label{tb:fields}
\end{table}
By using the tensor product of Eq.(\ref{tensorproduct}),
we can obtain the mass matrices of the up- and down-quarks 
in terms of the modular forms of weights $2$ and $4$, $Y^{S_3(2)}$ and $Y^{S_3(4)}$, as
\begin{equation}
	M_{u,d}=
	\left(
	\begin{array}{ccc}
	c^{u,d}+c'^{u,d}(Y_1(\tau')^2 -Y_2(\tau')^2) & 2c'^{u,d}Y_1(\tau')Y_2(\tau') & c^{u,d}_{13}Y_2({\tau'}) \\
	2c'^{u,d}Y_1(\tau')Y_2(\tau') & c^{u,d}-c'^{u,d}(Y_1(\tau')^2 -Y_2(\tau')^2) & -c^{u,d}_{13}Y_1(\tau') \\
	c^{u,d}_{31}Y_2(\tau') & -c^{u,d}_{31}Y_{1}(\tau') & c^{u,d}_{33}
	\end{array}\right),
	\label{massmatrix}
\end{equation}
where $c^{u,d}$ are expressed by the modular form of weight 4 and constant parameters $C^{u,d}$ as:
\begin{equation}
	c^{u,d}= C^{u,d} (Y_1(\tau')^2 +Y_2(\tau')^2)~.
\end{equation}
Hereafter we omit the superscript $S_3$ on $Y_i^{S_3}$.
It is noticed that the weight $4$ modular forms $Y^{(4)}$ appear in the first $2\times 2$ submatrix in Eq.(\ref{massmatrix})
while the weight $2$ modular forms $Y^{(2)}$ appear in  $(1,3)$, $(2,3)$, $(3,1)$ and $(3,2)$ elements of the mass matrix due to modular weights of quarks.
The modular forms do not appear in the $(3,3)$ entry.
The coefficients $c'^{u,d}$, $c^{u,d}_{13}$, $c^{u,d}_{31}$, $C^{u,d}$ and $c^{u,d}_{33}$ are arbitrary complex parameters,
where  $c^{u,d}_{33}$ can be taken to be real without loss of generality.
In addition to those parameters, the VEV of $\tau'$ is also a complex free parameter.

We have 20 free parameters enough to explain quark masses and CKM matrix elements.
Our quark mass matrices are completely consistent with the observed quark masses, the  CKM mixing angles and the CP violating phase.
In order to show how to reproduce the desired quark masses and mixing angles from the mass matrix of Eq.(\ref{massmatrix}), 
let us suppose the hierarchical structure of the mass matrix.
In the first $2\times 2$ submatrix, the mixing angle $\theta_{12}^{u,d}$ is
\begin{equation}
	\theta_{12}^{u,d}\simeq  
	\left |\frac{Y_1(\tau')Y_2(\tau')}{Y_1(\tau')^2 -Y_2(\tau')^2} \right |~,
\end{equation}
which is independent of $C^{u,d}$ and $c'^{u,d}$.
The ratio of  mixing angles $\theta_{23}^{u,d}$ and $\theta_{13}^{u,d}$ is given by the ratio of $(1,3)$ and $(2,3)$ entries of Eq.(\ref{massmatrix}) as:
\begin{equation}
	\frac{\theta_{13}^{u,d}}{\theta_{23}^{u,d}}\simeq 
	\left |\frac{Y_2(\tau')}{ Y_1(\tau')} \right |~,
\end{equation}
which is also independent of $c^{u,d}_{13}$.
It is remarked that the relations among mixing angles $\theta_{12}^{u,d}$, $\theta_{13}^{u,d}$ and $\theta_{23}^{u,d}$ are given only by the modulus $\tau'$.

In order to realize the quark mass hierarchy, we set parameters to suppress the $(1,1)$ entry, i.e.,
\begin{equation}
	c^{u,d}+c'^{u,d}(Y_1(\tau')^2 -Y_2(\tau')^2) \approx 0.
	\label{entry11}
\end{equation}
Moreover, we should put $|Y_1(\tau')^2|  \gg |Y_2(\tau')^2|$ to reproduce the hierarchical CKM mixing angles.
Then, we obtain 
\begin{equation}
	M_{u,d}\simeq
	\left(
	\begin{array}{ccc}
	0 & 2c'^{u,d}Y_1(\tau')Y_2(\tau') & c^{u,d}_{13}Y_2({\tau'}) \\
	2c'^{u,d}Y_1(\tau')Y_2(\tau') & -2c'^{u,d}Y_1(\tau')^2 & -c^{u,d}_{13}Y_1(\tau') \\
	c^{u,d}_{31}Y_2(\tau') & -c^{u,d}_{31}Y_{1}(\tau') & c^{u,d}_{33}
	\end{array}\right).
	\label{appromatrix}
\end{equation}
In order to obtain the Cabibbo angle, we take $|Y_2(\tau')/Y_1(\tau')| =\lambda\simeq 0.2$.
For this parameter, the quark mass matrices are written by 
\begin{equation}
	M_{u,d}\simeq 
	\left(
	\begin{array}{ccc}
	0 & \lambda\times 2c'^{u,d}Y_1(\tau')^2 & \lambda\times c^{u,d}_{13}Y_1({\tau'}) \\
	\lambda\times 2c'^{u,d}Y_1(\tau')^2 & -2c'^{u,d}Y_1(\tau')^2 & -c^{u,d}_{13}Y_1(\tau') \\
	\lambda \times c^{u,d}_{31}Y_1(\tau') & -c^{u,d}_{31}Y_{1}(\tau') & c^{u,d}_{33}
	\end{array}\right).
\end{equation}
The other angle $\theta_{23}^{u,d}$ can be estimated by setting $ c^{u,d}_{13}Y_1(\tau')/c^{u,d}_{33}$ properly, i.e. $ | c^{u,d}_{13}Y_1(\tau')/c^{u,d}_{33}| \sim \lambda^2$, which leads to observed $V_{cb}$.
On top of that, by fixing $c^{u,d}_{31}$ and $c^{u,d}_{13}$, we are able to fit second quark masses.
For the first quark masses, we need some tuning of parameters instead of Eq.(\ref{entry11}) as follows:
\begin{equation}
	c^{u,d}+c'^{u,d}(Y_1(\tau')^2 -Y_2(\tau')^2) = \varepsilon^{u,d}~ ,
	\label{eq:epsilon}
\end{equation}
with $\varepsilon^{u}$ and $\varepsilon^{d}$ being order of  $m_u$ and $m_d$, respectively.

Therefore, our quark mass matrices can be consistent with observed quark masses and CKM matrix by a specific set of parameters.
Indeed, we obtain $|Y_2(\tau')/Y_1(\tau')| \simeq 0.16$ and $ | c^{u(d)}_{13}Y_1(\tau')/c^{u(d)}_{33}| \simeq 0.06~(0.04)$ 
at a sample point, which is completely consistent with all experimental data, as shown in the next section.


\subsection{Numerical result of CKM mixing}
 
At first, we present a framework of our calculation of the CKM mixing angles $\theta_{12}$, $\theta_{23}$, $\theta_{13}$ and the  CP violating phase $\delta_{CP}$ which are expressed in the usual PDG convention \cite{Tanabashi:2018oca}.
Our numerical analysis is based on the mass matrices of Eq.\eqref{massmatrix}.
However, we use the relation of Eq.\eqref{eq:epsilon} and discuss $\varepsilon^{u,d}$ instead of $c^{u,d}$.
Since our mass matrices are given at a high energy scale such as the compactification scale,
we adopt the quark masses at GUT scale \cite{Antusch:2013jca, Bjorkeroth:2015ora} as input data to constrain the unknown parameters $c'^{u,d}$, $c^{u,d}_{13}$, $c^{u,d}_{31}$ and $\varepsilon^{u,d}$.
In the following calculations, we use $c^{u,d}_{33}=1$GeV unit.
The absolute values of those parameters are scanned around the values discussed in the previous section.
The phases of the parameters are scanned in $[-\pi,\pi]$.
We also scan parameters $\tau'$ in the complex plane by generating random numbers.

The scanned range of  ${\rm Im } [\tau']$ is $[0.5,3]$.
The lower-cut $0.5$ comes from the accuracy in calculating modular functions.
Indeed, the $q=e^{2\pi i\tau}$ expansions of modular functions are enough valid in ${\rm Im } [\tau'] \geq 0.5$ as seen in Eq. (\ref{expansionS3}).
The upper-cut $3$ is enough large for estimating  $Y_i$ in practice.
The modular function $Y_2$ decreases rapidly, $10^{-4}$ for ${\rm Im } [\tau'] = 3$ while $Y_1$ is almost $1/8$.
On the other hand,   ${\rm Re } [\tau]$ is scanned in the fundamental region of  $[-1, 1]$ 
because the modular function $Y_i(\tau')$ is given by $\eta(\tau'/2)$.
We calculate three CKM mixing angles and the CP violating phase in terms of the model parameters
while keeping the parameter sets leading to values allowed by the experimental data at $3\,\sigma$ C.L..

We use the following quark Yukawa couplings in order to constrain the model parameters at the GUT scale $2\times 10^{16}$\,GeV where $\tan\beta=10$ is taken:
\cite{Antusch:2013jca, Bjorkeroth:2015ora}:
\begin{align}
	\begin{aligned}
	&y_d=(4.84\pm 1.07) \times 10^{-6}, \quad y_s=(9.59\pm 1.04) \times 10^{-5}, \quad y_b=(7.01\pm 0.178) \times 10^{-3}, \\
	\rule[15pt]{0pt}{1pt}
	&y_u=(2.88\pm 1.79) \times 10^{-6}, \quad y_c=(1.41{\pm 0.0987}) \times 10^{-3}, \quad y_t=0.520\pm 0.0315  ~~,
	\end{aligned}\label{yukawa}
\end{align}
which give quark masses as $m_q=y_q v_H$ with $v_H=174.104$ GeV.
We also use the following CKM mixing angles and the CP violating phase to focus on parameter regions consistent with the 
experimental data \cite{Antusch:2013jca, Bjorkeroth:2015ora}:
\begin{align}
	\begin{aligned}
	&\theta_{12}=13.027^\circ\pm 0.0814^\circ ~ , \qquad \qquad
	\theta_{23}=2.054^\circ\pm 0.384^\circ ~ , \quad  \\
	\rule[15pt]{0pt}{1pt}
	&\theta_{13}=0.1802^\circ\pm 0.0281^\circ~ , \qquad \qquad
	\delta_{CP}=69.21^\circ\pm 6.19^\circ~ .
	\end{aligned}\label{CKM}
\end{align}
The error widths in Eqs.(\ref{yukawa}) and (\ref{CKM}) represent $1\sigma$ interval.
In our numerical calculation, we use $1\sigma$ interval for quark masses and present favorable regions which are consistent with experimental data of the CKM matrix with $3\sigma$ interval.

We show one numerical sample point which is completely consistent with quark masses and CKM elements:
\begin{align}
	\begin{aligned}
	&{\rm Re } [\tau'] = -0.4216~ ,\qquad\qquad\qquad\qquad  {\rm Im }[\tau']=1.4261 ~ , \\
	&\varepsilon^{u}=(1.172-i~ 8.3547)\times 10^{-6}~, \qquad c'^{u}=(0.9469-i~ 8.3220)\times 10^{-3},  \\
	&c^{u}_{13}=-0.1858+i~ 0.5003~, \qquad\qquad ~~ c^{u}_{31}=-0.2516-i~ 0.1697~, \\
	&\varepsilon^{d}=(0.2530+i~ 1.1190)\times 10^{-3}~, \qquad c'^{d}=0.1069+i~ 0.4273,  \\
	&c^{d}_{13}=-0.1707+i~ 0.3066~, \qquad \qquad\quad c^{d}_{31}=-0.0715-i~ 0.2165~,
	\end{aligned}
\end{align}
in $c^{u,d}_{33}=1$GeV units.
Then, the calculated CKM mixing angles and the CP violating phase are:
\begin{align}
	\begin{aligned}
	& \theta_{12}=12.99^\circ  ~ , \qquad \qquad 
	 \theta_{23}=1.31^\circ ~ , \\
	\rule[15pt]{0pt}{1pt}
	& \theta_{13}=0.20^\circ~ , \qquad \qquad~ 
	\delta_{CP}=65.17^\circ~ ,  
	\end{aligned}\label{sample}
\end{align}
which fit to the data in Eq.(\ref{CKM}).
We also show our successful results in Figs.1 and 2.
\begin{figure}[t!]
	\begin{tabular}{ccc}
		\begin{minipage}{0.475\hsize}
			\includegraphics[width=\linewidth]{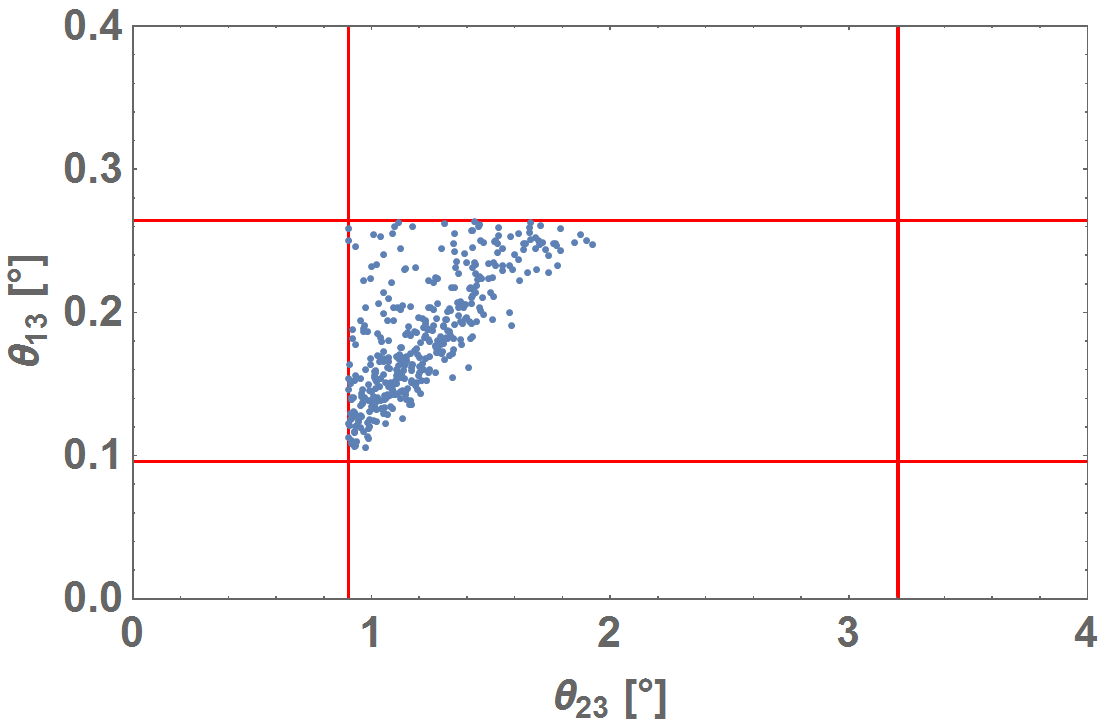}
			\caption{The prediction of $\theta_{13}$ versus  $\theta_{23}$.
				The horizontal and vertical red lines represent the upper and lower bounds of the experimental data with $3 \ \sigma$ at GUT scale.}
		\end{minipage}
		\phantom{=}
		\begin{minipage}{0.475\linewidth}
			\includegraphics[width=\linewidth]{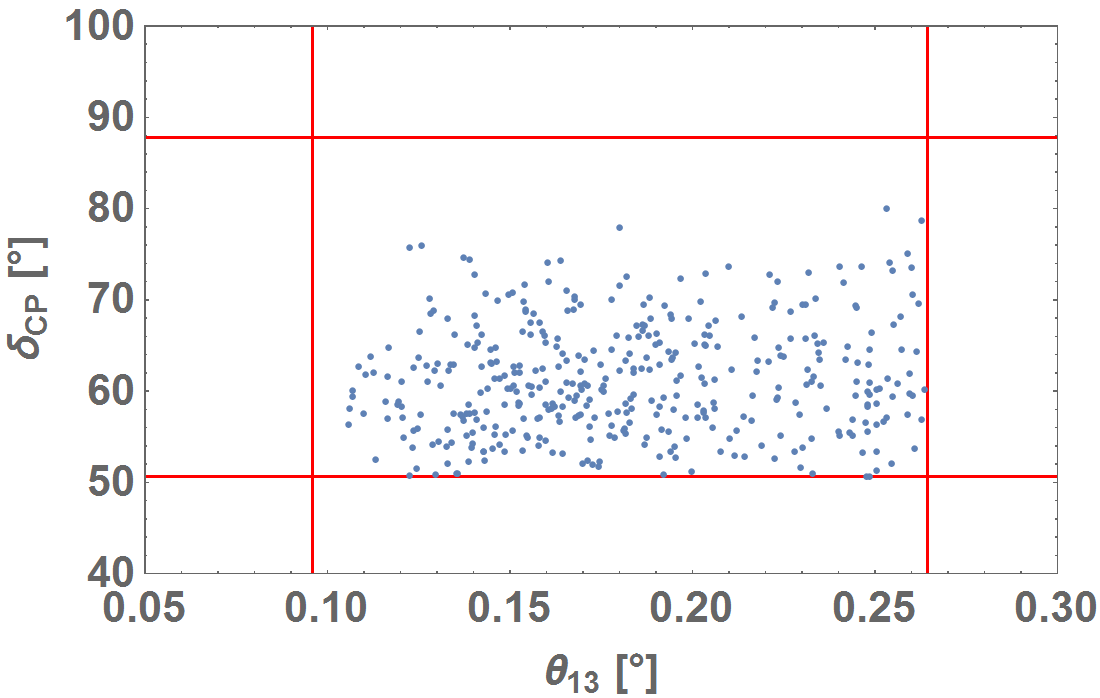}
			\caption{The prediction of $\delta_{CP}$ versus $\theta_{13}$.
				The horizontal and vertical red lines represent the upper and lower bounds of the experimental data with $3 \ \sigma$ at GUT scale.}
		\end{minipage}
	\end{tabular}
\end{figure}
In Fig.1, we plot the predicted correlation of  $\theta_{23}$ and $\theta_{13}$,
where $\theta_{23} > 2^\circ $ is not allowed.
In Fig.2, we plot the predicted  $\delta_{CP}$
versus $\theta_{13}$, where $\delta_{CP}>80^\circ $ is excluded.
The different densities of points have no physical meaning in Figs.\,1 and 2.

We suppose the hierarchical structure of the quark mass matrix to obtain above numerical result.
In order to get strong predictions in the quark sector, we need further analyses taking account of possible GUT parameters such as the mass scale and the intermediate mass spectra.
However, it can be stated that our quark mass matrices are consistent with the experimental data.
This situation is occurred by the characteristic flavor structure of $(1,1)$,  $(1,2)$, $(2,2)$, $(1,3)$, and  $(2,3)$ entries of quark mass matrices in Eq.(\ref{appromatrix}), which comes from the modular $S_3$ symmetry.


\section{Baryon and lepton number violations}

Here, we study baryon and lepton number violations in our model.
Baryon and lepton numbers are violated in a generic supersymmetric standard model.
That leads to a fast proton decay if the superpartner mass scale is sufficiently low, e.g. a TeV scale.
R-parity is often assumed not to lead to a fast proton decay.
(See for reviews on R-parity, e.g. \cite{Dreiner:1997uz,Barbier:2004ez}.)
Quarks, leptons and Higgs scalar fields have even R-parity, while their superpartners have odd R-parity.
Such an R-parity allows Yukawa coupling terms for quarks and leptons in the superpotential as well as the Higgs $\mu$-term.
Apart from these R-parity conserving terms, the following terms:
\begin{equation}
	W_{non-R} = m_i L_i H_u + \lambda_{ijk}L_iL_je_k + \lambda'_{ijk}L_iQ_j d_k + \lambda''_{ijk} u_i d_j d_k,
\end{equation}
are possible in the renormalizable superpotential.
These terms violate baryon and lepton numbers.
For example, the combination of the third and fourth terms leads to a fast proton decay.
Thus, these couplings are strongly constrained as \cite{Barbier:2004ez}
\begin{equation}
	\label{eq:proton-decay}
	\lambda'_{11k}\lambda''_{11k} < 7\cdot10^{-26}\left( \frac{\tilde m_{dk}}{1 {\rm TeV}}\right)^2,
\end{equation}
where $\tilde m_{dk}$ denotes the $k$-generation of down-sector squark mass 
because of the proton life time $\tau (P \rightarrow e \pi) > 10^{33} yr$.
R-parity forbids $W_{non-R}$, and a fast proton decay does not occur 
even if the squark mass scale is low.

Also, there are dimension five operators violating baryon and/or lepton numbers in 
the superpotential, 
\begin{equation}
	\label{eq:dim-5}
	QQQL, \quad uude, \quad QQQH_d, \quad QueH_d, \quad LH_uLH_u, \quad L H_uH_d H_u.
\end{equation}
Some of them are forbidden by R-parity, while R-parity allows $QQQL$, $uude$ and $LH_uLH_u$.

We study baryon and lepton number violations in our model without imposing R-parity.
Recall our assignment of modular weights.
The first and second generations of quarks have the modular weight $-2$, while 
the third generation of quarks have the vanishing modular weight.
All of the leptons have the modular weight $-1$, and both Higgs fields have 
the vanishing modular weight.
In our model, all of the couplings as well as masses must be modular forms of even weights.
This setup leads to the $Z_2$ symmetry, where all of the leptons are $Z_2$ odd 
and quarks and Higgs fields are $Z_2$ even.
That is, the $Z_2$ lepton number.
This symmetry forbids most of the terms in $W_{non-R}$.
Only the terms $ \lambda''_{ijk} u_i d_j d_k$ are allowed.
Because of $\lambda' =0$, the proton decay constraint (\ref{eq:proton-decay}) is satisfied 
for any value of $\lambda''$.
Similarly, some dimension five operators in Eq.(\ref{eq:dim-5}) are forbidden, 
but only $QQQH_d$ and  $LH_uLH_u$ are allowed.
That is, the Weinberg operator is allowed.
That is consistent with the fact that the seesaw mechanism works and the mass term 
(\ref{eq:seesaw-mass}) is written.
In addition to this mass term, another Weinberg operator suppressed by the compactification scale $M_c$, 
i.e. $LH_uLH_u/M_c$ can be generated in superstring theory.
It is expected that $M_c$ is much higher than $\Lambda$, and typically
$M_c$ is of ${\cal O} (10^{16} - 10^{18})$ GeV, although lower compactification scale would be possible.
For $\Lambda \ll M_c$, the above mass term $LH_uLH_u/M_c$ is negligible.
When $M_c$ is comparable to $\Lambda$, our results are the same by changing values of our parameters.


For concreteness,  we explicitly write the possible terms of 
$ \lambda''_{ijk} u_i d_j d_k$, 
\begin{align}
\begin{aligned}
&	a~\epsilon^{pqr}\begin{bmatrix}
\begin{pmatrix}
Y_1(\tau')(Y_1(\tau')^2+Y_2(\tau')^2) \\ Y_2(\tau')(Y_1(\tau')^2+Y_2(\tau')^2)
\end{pmatrix}_\mathbf{2}
\times
\begin{pmatrix}
u_1^{p} \\ u_2^{p}
\end{pmatrix}_\mathbf{2}
\end{bmatrix}_\mathbf{1'}
\begin{bmatrix}
\begin{pmatrix}
d_1^{q} \\ d_2^{q}
\end{pmatrix}_\mathbf{2}
\times
\begin{pmatrix}
d_1^{r} \\ d_2^{r}
\end{pmatrix}_\mathbf{2}
\end{bmatrix}_\mathbf{1'}  \\ 
+~ &b~(Y_1(\tau')^2+Y_2(\tau')^2)
\epsilon^{pqr}
\begin{bmatrix}
\begin{pmatrix}
u_1^{p} \\ u_2^{p}
\end{pmatrix}_\mathbf{2}
\times
\begin{pmatrix}
d_1^{q} \\ d_2^{q}
\end{pmatrix}_\mathbf{2}
\end{bmatrix}_\mathbf{1'}
d_3^{r} \\
+~& c~\epsilon^{pqr}
\begin{bmatrix}
\begin{pmatrix}
Y_1(\tau')^2-Y_2(\tau')^2 \\ -2Y_1(\tau')Y_2(\tau')
\end{pmatrix}_\mathbf{2}
\times
\begin{bmatrix}
\begin{pmatrix}
u_1^{p} \\ u_2^{p}
\end{pmatrix}_\mathbf{2}
\times
\begin{pmatrix}
d_1^{q} \\ d_2^{q}
\end{pmatrix}_\mathbf{2}
\end{bmatrix}_\mathbf{2}
\end{bmatrix}_\mathbf{1'}
d_3^{r} \\
+~&d~\epsilon^{pqr}u_3^{p}d_3^{q}
\begin{bmatrix}
\begin{pmatrix}
d_1^{r} \\ d_2^{r}
\end{pmatrix}_\mathbf{2}
\times
\begin{pmatrix}
Y_1(\tau') \\ Y_2(\tau')
\end{pmatrix}_\mathbf{2}
\end{bmatrix}_\mathbf{1} \\
+~& e~\epsilon^{pqr}(Y_1(\tau')^2+Y_2(\tau')^2) u_3^{p}
\begin{bmatrix}
\begin{pmatrix}
{d}_1^{q} \\ {d}_2^{q}
\end{pmatrix}_\mathbf{2}
\times
\begin{pmatrix}
{d}_1^{r} \\ {d}_2^{r}
\end{pmatrix}_\mathbf{2}
\end{bmatrix}_\mathbf{1'}
\end{aligned}\label{eq:coupling}
\end{align}
where $a$, $b$, $c$, $d$ and $e$ are arbitrary coefficients.
Therefore, $\lambda''_{ijk}$ are written by 
\begin{align}
\begin{aligned}
\lambda''_{212} &= -\lambda''_{221}=aY_1(\tau')(Y_1(\tau')^2 + Y_2(\tau')^2), \\
\lambda''_{121} &= -\lambda''_{112}=aY_2(\tau')(Y_1(\tau')^2 + Y_2(\tau')^2), \\
\lambda''_{123} &= (b-c)Y_1(\tau')^2+ (b+c)Y_2(\tau')^2, \\
\lambda''_{213} &= -(b+c)Y_1(\tau')^2+ (c-b)Y_2(\tau')^2, \\
\lambda''_{113} &= -\lambda''_{223} = 2cY_1(\tau')Y_2(\tau'),  \\
\lambda''_{331} &= dY_1(\tau'), \\
\lambda''_{332} &= dY_2(\tau'), \\
\lambda_{312}^{\prime\prime} &=-\lambda''_{321}=e(Y_1(\tau')^2+Y_2(\tau')^2).
\end{aligned}\label{eq:lambda''}
\end{align}
These couplings are important from the viewpoints of flavor changing processes and CP violation 
phenomena \cite{Barbier:2004ez}.

The modular symmetry can be anomalous
\cite{Derendinger:1991hq,Ibanez:1992hc,Kobayashi:2016ovu}.\footnote{See also \cite{Araki:2007ss}.}
Because of the transformation (\ref{eq:modular-chiral}), coefficients of mixed modular symmetry anomalies with 
the gauge group $G$ and gravity are written by 
$\sum_I k_IT(R_I)$ and $\sum_I k_I$, respectively, where 
$T(R_I)$ denotes the Dynkin index for representation $R_I$ of the chiral matter $\phi^{(I)}$ under $G$. 
In supergravity theory, there are other contributions \cite{Derendinger:1991hq,Ibanez:1992hc}.
Such anomalies can be canceled by Green-Schwarz mechanism within the 
framework of superstring theory.\footnote{These anomalies lead to phenomenologically interesting aspects \cite{Ibanez:1991zv,Kawabe:1994mj}.}
The modular symmetry is anomalous in our model.
We assume that such anomalies can be canceled by the Green-Schwarz mechanism.
Suppose that the gauge kinetic function $f$ is written by a single field $S$, 
which is  modulus or dilaton.
The real part of $\langle f \rangle = \langle S \rangle$ determines the gauge coupling,
$g^{-2 }= \langle \mbox{Re}[f] \rangle = \langle \mbox{Re}[S] \rangle$.
The Green-Schwarz mechanism implies that the field $S$ shifts under anomalous symmetry.
That is, under the modular transformation the field $S$ transforms  
\begin{equation}
	S \rightarrow S + \frac{\delta_{GS}}{8\pi^2}\ln(c\tau + d),
\end{equation}
where $\delta_{GS}$ is anomaly coefficient and of ${\cal O}(1-10)$.
This means that $e^{-aS}$ behaves as if it had a modular weight $a\delta_{GS}/(8\pi^2)$.
In addition, non-perturbative effects such as D-brane instanton effects and gaugino 
condensation would induce terms such as $e^{-aS} \prod_i\Phi_i$ in the superpotential, where $\Phi_i$ are chiral superfields.
Such non-perturbative terms must be invariant under anomalous symmetries including 
the shift of $S$.
Thus, the above $Z_2$ lepton number is violated by non-perturbative effects, $e^{-aS}$.
That would induce dangerous terms, e.g. $\lambda'_{ijk}L_iQ_j d_k$, 
where $\lambda'_{ijk} \propto e^{-aS}$.
Here, $e^{-aS}$ must be modular weight odd so that such terms are induced by non-perturbative effects.
Thus, the minimum one is $a=8\pi^2/\delta_{GS}$.
Then, we find $\lambda'_{ijk} \sim e^{-8\pi^2/(\delta_{GS}g^2)} =  e^{-2\pi \alpha^{-1}/\delta_{GS}}$, 
up to a coefficient, where $\alpha = g^2/(4 \pi)$.
For example, for $\alpha^{-1}=25$ we obtain 
\begin{equation}
	e^{-2\pi \alpha^{-1}/\delta_{GS}}  \sim 10^{-68}, 10^{-34}, 10^{-23}, 10^{-17}, 10^{-14}, 
\end{equation}
for $\delta_{GS} = 1,2,3,4,5$ , respectively. 
Superpartner masses would be required to be heavy.
A large value of $\alpha^{-1}/\delta_{GS}$ would be favorable to satisfy the proton decay constraint 
(\ref{eq:proton-decay}).
That is, larger value of $\alpha^{-1}$ might be favorable.
Note that $\lambda''$ in Eq.(\ref{eq:lambda''}) would also include some suppression factors.


\section{Summary}

We have studied a flavor model where the quark sector has the $S_3$ modular symmetry
while the lepton sector has the  $A_4$  modular symmetry.
The masses and mixing angles of the lepton sector is reproduced in the flavor symmetry by using the finite modular group $\Gamma_3 \simeq A_4$.
In this work, we have proposed quark mass matrices in the flavor model of the finite modular group $\Gamma_2 \simeq S_3$,
where the first two families are assigned to $\bf 2$ while the third family to non-trivial singlet $\bf 1'$.
Both weight $2$ and $4$ modular forms are available in our framework.
Then, there appears the characteristic flavor structure in $(1,1)$, $(1,2)$, $(2,2)$, $(1,3)$, and $(2,3)$ entries of 
the quark mass matrix of Eq.(\ref{massmatrix}). 
Supposing the hierarchical structure of the quark mass matrix,
we have obtained desirable quark mass matrices, which are completely consistent with the observed quark masses, CKM mixing angles and the CP violating phase.


Our setup, the quark and lepton sectors have different flavor symmetries, 
has interesting implications from the viewpoint of baryon and lepton number violations.
At the perturbative level, a fast proton decay can be forbidden even for light superpartners.
However, non-perturbative effects can break such a situation and a larger $\alpha^{-1}$ is favorable.

We have assumed that the VEVs of the moduli, $\tau$ and $\tau'$, are free parameters.
It is very important to know how to fix their VEVs at favorable values to realize 
fermion masses and mixing angles in our scenario.
Fixing their VEVs is the so-called moduli stabilization problem which is one of the most important issues in superstring theory \cite{Kachru:2003aw,Balasubramanian:2005zx}.
It is beyond our scope.

\vspace{0.5cm}
\noindent

{\bf Acknowledgement}  

This work is supported by  MEXT KAKENHI Grant Number JP17H05395 (TK), and 
JSPS Grants-in-Aid for Scientific Research 
16J05332 (YS), 15K05045 (MT), and 18J11233 (THT).


\appendix

\section*{Appendix}

\section{Multiplication rule of $A_4$ group}
\label{sec:multiplication-rule}
We use the multiplication rule of the $A_4$ triplet as follows:
\begin{align}
\begin{pmatrix}
a_1\\
a_2\\
a_3
\end{pmatrix}_{\bf 3}
\otimes 
\begin{pmatrix}
b_1\\
b_2\\
b_3
\end{pmatrix}_{\bf 3}
&=\left (a_1b_1+a_2b_3+a_3b_2\right )_{\bf 1} 
\oplus \left (a_3b_3+a_1b_2+a_2b_1\right )_{{\bf 1}'} \nonumber \\
& \oplus \left (a_2b_2+a_1b_3+a_3b_1\right )_{{\bf 1}''} \nonumber \\
&\oplus \frac13
\begin{pmatrix}
2a_1b_1-a_2b_3-a_3b_2 \\
2a_3b_3-a_1b_2-a_2b_1 \\
2a_2b_2-a_1b_3-a_3b_1
\end{pmatrix}_{{\bf 3}}
\oplus \frac12
\begin{pmatrix}
a_2b_3-a_3b_2 \\
a_1b_2-a_2b_1 \\
a_3b_1-a_1b_3
\end{pmatrix}_{{\bf 3}\  } \ , \nonumber \\
\nonumber \\
{\bf 1} \otimes {\bf 1} = {\bf 1} \ , \qquad &
{\bf 1'} \otimes {\bf 1'} = {\bf 1''} \ , \qquad
{\bf 1''} \otimes {\bf 1''} = {\bf 1'} \ , \qquad
{\bf 1'} \otimes {\bf 1''} = {\bf 1} \  .
\end{align}
More details are shown in the review~\cite{Ishimori:2010au,Ishimori:2012zz}.

\section{Tensor products of modular forms}

Here we give tensor products of modular forms of weight 2 correspond to 
the $S_3$ doublet.
We can write product of $(Y_1(\tau'),  Y_2(\tau'))^T$
\begin{eqnarray}
\begin{pmatrix}
Y_1(\tau') \\ Y_2(\tau')
\end{pmatrix}_\mathbf{2}
\times
\begin{pmatrix}
Y_1(\tau') \\ Y_2(\tau')
\end{pmatrix}_\mathbf{2}
\times
\begin{pmatrix}
Y_1(\tau') \\ Y_2(\tau')
\end{pmatrix}_\mathbf{2}
&=&
\begin{pmatrix}
Y_1(\tau') \\ Y_2(\tau')
\end{pmatrix}_\mathbf{2}
\times
\left(Y_1(\tau') ^2 +Y_2(\tau')^2\right)_\mathbf{1} \notag \\
&+&
\begin{pmatrix}
Y_1(\tau') \\ Y_2(\tau')
\end{pmatrix}_\mathbf{2}
\times
\begin{pmatrix}
Y_1(\tau') ^2 - Y_2(\tau')^2 \\ -2Y_1(\tau')Y_2(\tau')
\end{pmatrix}_\mathbf{2} \nonumber \\
&=&
\left[Y_1(\tau') \left(Y_1(\tau') ^2 - 3Y_2(\tau')^2\right) \right]_\mathbf{1} \nonumber \\
&+&
\left[Y_2(\tau') \left(Y_2(\tau') ^2 - 3Y_1(\tau')^2\right) \right]_\mathbf{1'} \notag \\
&+&
2\begin{pmatrix}
Y_1(\tau')(Y_1(\tau')^2+Y_2(\tau')^2) \\ Y_2(\tau')(Y_1(\tau')^2+Y_2(\tau')^2)
\end{pmatrix}_\mathbf{2} .
\end{eqnarray}



\end{document}